\documentclass[aps,prb,twocolumn,showpacs]{revtex4}

\usepackage{mathptm}    
\usepackage{dcolumn}                    
\usepackage{bm}                         
\usepackage{graphicx}

\usepackage{times}

\begin{document}

\title{Theory of Neutron Scattering in High-T$_c$ Cuprates: Two Component Spin-Fermion Model}

\author{Yunkyu Bang}
\affiliation{Department of Physics, Chonnam National University,
Kwangju 500-757, and Asia Pacific Center for Theoretical Physics, Pohang 790-784, Korea}

\begin{abstract}
Recent neutron scattering experiments have revealed that the
generic form of the magnetic excitations in the high-Tc cuprates
of wide range of doping has the so-called "hourglass" shape; it
features both upward and downward excitations at the
incommensurate (IC) momenta spanning from the resonance peak at
the commensurate momentum $(\pi,\pi)$. We propose the
two-component spin-fermion model as a minimal phenomenological
model which has both local spins and itinerant fermions as
independent degrees of freedom. Our calculations of the dynamic
spin correlation function provide good agreement with experiments
and show: (1) the upward dispersion branch of magnetic excitations
is mostly due to the local spin excitations; (2) the downward
dispersion branch is from collective spin excitations of fermions;
(3) the resonance mode is a mixture of both degrees of freedom.
\end{abstract}


\date{\today}
\maketitle

The study of spin dynamics has been a key research interest since
the discovery of the high-T$_c$ superconductor because it is
expected that the spin correlation holds crucial information for
the mechanism of the high-T$_c$ superconductivity (HTSC). For long
the two main observations in the neutron scattering experiments of
high-T$_c$ cuprates are (1) the incommensurate (IC) peaks  at low
energy or at quasielastic excitations \cite{Cheong,Dai} and (2)
the so-called resonance peak at commensurate wave vector at
relatively high energy (30 $\sim$ 50 meV) \cite{resonance}. In
early experiments the IC peaks were observed only in the deeply
underdoped lanthanum cuprate (LCO) and the resonance mode was the
hallmark of the fully doped two layer yttrium cuprate (YBCO).
However, later experiments reveal that both features appear in
both cuprates, although sensitively depending on doping level. And
with more recent
experiemnts\cite{Arai99,Hayden04,Tranquada04,Pailhes04,Hinkov04,Keimer-condmat06},
a unifying form of the magnetic excitations in the cuprate
superconductors has emerged as "hourglass" shape of excitations
around the wave vector (1/2,1/2) (hereafter in units of $2 \pi /
a$), in which the low energy IC excitations form the downward
dispersion branch and the high energy IC excitations form the
upward dispersion branch, and the two branches of excitations
merge at the commensurate momentum and at the resonance frequency.

It is a pressing question to understand the origin  of this
"hourglass" shape excitations. Theoretical proposals up to now can
be classified into two groups: (1) theories based on the spin
dynamics in the presence of stripes \cite{stripes,stripes-2}, and
(2) Fermi liquid type theories of itinerant fermions
\cite{FL-theories, FL-theories-2}.
The key idea of the first group of theories is that the stripes
formed by doping in the two dimensional Cu-O plane can split the
commensurate spin wave excitations into two IC branches at the
wave vectors (1/2 $+ \epsilon$, 1/2) and (1/2 $- \epsilon$ , 1/2 )
or at their symmetry rotated vector positions by $ x
\longleftrightarrow y$ depending on the directions of the stripes.
The dispersions from each branch of the two IC modulation cross at
the commensurate wave vector (1/2,1/2) at a higher energy, which
can be identified as the resonance mode. This picture provides a
qualitative explanation to the hourglass dispersion and the
resonance mode. However, this type of theories has difficulty to
be extended to the higher doping regime where the presence and the
nature of the stripes is questionable.
The second group of proposals are itinerant fermion theories with
interaction \cite{FL-theories, FL-theories-2}. In this type of
theories, the resonance mode and the downward dispersion can be
obtained, but the upward dispersion branch is not yet
satisfactorily reproduced.

In this paper, we propose a two component spin-fermion model
\cite{Bang} as a minimal phenomenological model to provide a
natural and unifying explanation for the above mentioned neutron
experiments of high-T$_c$ cuprates. In this phenomenological
model, the minimal set of low energy degrees of freedom are the
spin wave excitations of local spins and the continuum
particle-hole excitations of fermions.
A similar phenomenological theory is also known as one component
spin-fermion model and has been intensively studied by Pines and
coworkers.\cite{Pines-review} The key difference from the one
component model, is the presence of the spin wave excitations
directly from the local spins in addition to the usual collective
spin density excitations from fermions. Despite a gross similarity
to the one component model, we argue that the necessity of the
local spins for the minimal model of HTSC is impelling from the
neutron experiments. In a mixed momentum and real-space
representation the corresponding Hamiltonian is written as

\begin{equation}
 H = \sum_{{\bf k}, \alpha} c^{\dag}_\alpha({\bf
 k})\varepsilon({\bf k})c_\alpha({\bf
 k}) + \sum_{{\bf r},\alpha, \beta} g {\bf \vec{S}}({\bf r}) \cdot
c^{\dag}_\alpha({\bf r}){\bf \vec{\sigma}}_{\alpha \beta}c_\beta({\bf
 r}) + H_S,
\end{equation}

\noindent where the first term is the fermionic kinetic energy and
the second term describes the coupling between local spins  ${\bf
\vec{S}}({\bf r})$ and the spin density of the conduction
electrons. The last term  represents an effective low-energy
Hamiltonian for the local spins. When the local spins have a short
range AFM correlation, the bare (before coupling to the fermions)
spin correlation function has the general form as
follows\cite{AFM}.
\begin{equation}
\chi_{0,S} ^{-1} ({\bf q}, \Omega) = \chi_{0,S} ^{-1}({\bf Q}, 0)
\cdot [1 +  \xi^{2} |{\bf q} -{\bf Q}|^2 - \Omega^2 / \Delta_{SG}
^2 ],
\end{equation}
\noindent where ${\bf Q}$ the 2D AFM ordering vector, and the spin
gap energy $\Delta_{SG}$ and the magnetic correlation length $\xi$
combine to give the spin wave velocity $v_s = \Delta_{SG} \cdot
\xi $. Counting the coupling term to one loop order, the dressed
spin correlation functions of the model are written as follows.

\begin{eqnarray}
\chi^{-1} _{S} ({\bf q},\Omega) &=& \chi^{-1} _{0,S} ({\bf
q},\Omega) - g^2 \cdot \chi_{0,f} ({\bf q},\Omega) \\
\chi^{-1} _{f} ({\bf q},\Omega) &=& \chi^{-1} _{0,f} ({\bf
q},\Omega) - g^2 \cdot \chi_{0,S} ({\bf q},\Omega)
\end{eqnarray}

\noindent where $\chi_{0,f}$ is the noninteracting spin
susceptibility of the conduction band of the fermions and
$\chi_{0,S}$ is introduced in Eq.(2).

Having two degrees of freedom in the model, two spin
susceptibilities $\chi_{S}$ and $\chi_{f}$ should be calculated on
equal footing. Previous studies of the local spin correlation
embedded in the fermion bath \cite{AFM} considered only the
imaginary part of $\chi_{0,f}$ ( Landau damping ) to damp the spin
wave excitations in Eq.(3) and the real part of $\chi_{0,f}$ is
assumed either already included in the local spin dynamics
described in Eq.(2)or its effect unimportant. In fact, when the
coupling $g$ is weak, this approach is reasonable. But in the
strong coupling limit (when the dimensionless coupling constant
$\lambda\equiv g^2 \cdot \chi_{0,f}({\bf Q},0) \cdot
\chi_{0,S}({\bf Q},0) \sim O(1)$), including both the real and
imaginary parts as in the above equations is crucial. In the
strong coupling limit, both spin susceptibilities $\chi_{S} ({\bf
q},\Omega)$ and $\chi_{f} ({\bf q},\Omega)$ become a mixture of
both the local spins and the itinerant fermions and they
assimilate to each other with increasing the coupling strength
$\lambda$.

To make a contact with experiments, we use a tight binding model
for the fermion dispersion

\begin{equation}
\epsilon ({\bf k}) = -2 t (\cos(k_x) + \cos(k_y)) - 2 t^{'}
\cos(k_x) \cdot \cos(k_y)-\mu.
\end{equation}

\noindent For calculations in this paper, we chose t$^{'}$=-0.4 t,
and $\mu= -0.81t$. The energy scale $t$ and the choice of
parameters $t^{'}$, $\mu$  will be discussed later with the
numerical results. The fermion susceptibility $\chi_{0,f}$ is
calculated both in normal state (NS) and in superconducting state
(SS) assuming a canonical d-wave pairing $\Delta ({\bf k}) =
\Delta_0 [ \cos(k_x) -\cos(k_y)]$.

Fig.1(a) shows $Im \chi_{f} ({\bf q},\Omega)$ scanned along ${\bf
q}= (h,1/2)$ in the superconducting state. The superconducting gap
$\Delta_0 =0.2 t$ and the bare spin gap $\Delta_{SG}=1.1 t$ is
chosen; the physical spin gap is strongly renormalized by coupling
with fermions. The dimensionless coupling constant $\lambda=0.80$
is chosen. The main effect of the coupling is to pull down the
bare spin gap $\Delta_{SG}$ below the particle-hole excitation gap
of $\chi_{0,f} ({\bf q},\Omega)$ ($\sim 2 \Delta_0$), which then
forms a sharp resonance peak at ${\bf Q}=(1/2,1/2)$. Centering
from this resonance mode, both the downward dispersion branch and
the upward dispersion branch span out. The origin of the upward
dispersion is apparently from the local spin wave mode (see
Eq.(2)) and the origin of the downward dispersion is the itinerant
spin excitations of $\chi_{0,f}$. This fact is identified in
Fig.1(b) which shows $Im \chi_{0,f} ({\bf q},\Omega)$ scanned
along ${\bf q}= (h,1/2)$ in the superconducting state. The shape
and strength of the downward whisker like excitations in $Im
\chi_{0,f}$ is quite sensitive to the band structure (controlled
by $t, t^{'}$, etc), Fermi surface curvature (controlled by $\mu$,
$t^{'}$), and the size of the d-wave gap $\Delta({\bf k})$.

With the coupling strength $\lambda=0.8$, the dressed
susceptibility $\chi_{f} ({\bf q},\Omega)$ obtains features of
both the local spin susceptibility $\chi_{0,S}$ and the itinerant
spin susceptibility $\chi_{0,f}$, and the behavior of the dressed
susceptibility $\chi_{S} ({\bf q},\Omega)$ is qualitatively
similar to the one of $\chi_{f} ({\bf q},\Omega)$ in this coupling
strength. Therefore we show only the results of $\chi_{f}({\bf
q},\Omega)$ in Fig.1.; the experimental data of neutron scattering
should be the contributions from both susceptibilities. With a
smaller coupling strength ($\lambda < 0.5$) the two spin
susceptibilities retain more of their original characteristics of
the local spin wave and the itinerant fermion susceptibility,
respectively.

\begin{figure}
\noindent
\includegraphics[width=100mm]{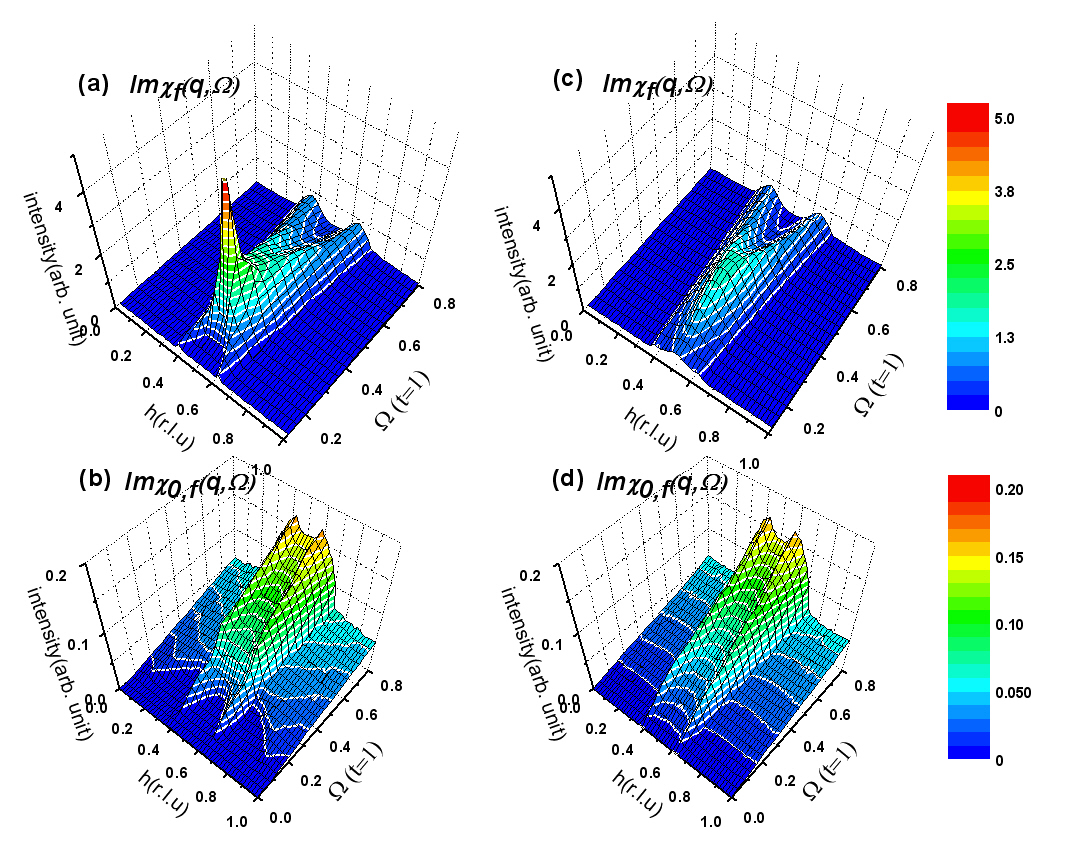}
\caption{(a) The dressed spin susceptibility $Im \chi_{f}({\bf
q}=(h,1/2), \Omega)$ in superconducting state. Parameters are
$\Delta_{SG}=1.1 t$, $\Delta_0 = 0.2 t$, and $\lambda=0.8$ (b) The
bare spin susceptibility $Im \chi_{0,f}({\bf q}, \Omega)$ in
superconducting state. (c) $Im \chi_{f}({\bf q}, \Omega)$ in
normal state ($\Delta_0 =0$). (d) $Im \chi_{0,f}({\bf q}, \Omega)$
in normal state.} \label{fig1}
\end{figure}

Fig.1(c,d) are the same plots as in Fig.1(a,b) but in normal
state. First, the resonance peak becomes a completely overdamped
mode having only a hump like structure in $Im \chi_{f} ({\bf
q},\Omega)$. Second, the downward whisker like dispersion
disappears because the free fermion susceptibility $\chi_{0,f}
({\bf q},\Omega)$ in normal state has no such structure as seen in
Fig.1(d). Lastly, the upward dispersion remains almost similar to
the case of the superconducting state. In experiments the upward
dispersion in normal state appears more smeared
\cite{Keimer-condmat06}. This could be due to the effect of
enhanced damping process at higher temperatures which is not
included in our calculations.
The results of Fig.1(a,c) successfully reproduce the main features
of recent neutron scattering experiments in high-T$_c$ cuprates
\cite{Arai99,Hayden04,Tranquada04,Pailhes04, Hinkov04,
Keimer-condmat06}, ie., the resonance mode in superconducting
state, the hourglass shape of the upward and downward dispersions,
and their drastic change in normal and superconducting states. In
particular, these results are in excellent agreement with the data
of YBCO$_{6.6}$ \cite{Keimer-condmat06}.

\begin{figure}
\noindent
\includegraphics[width=95mm,height=75mm]{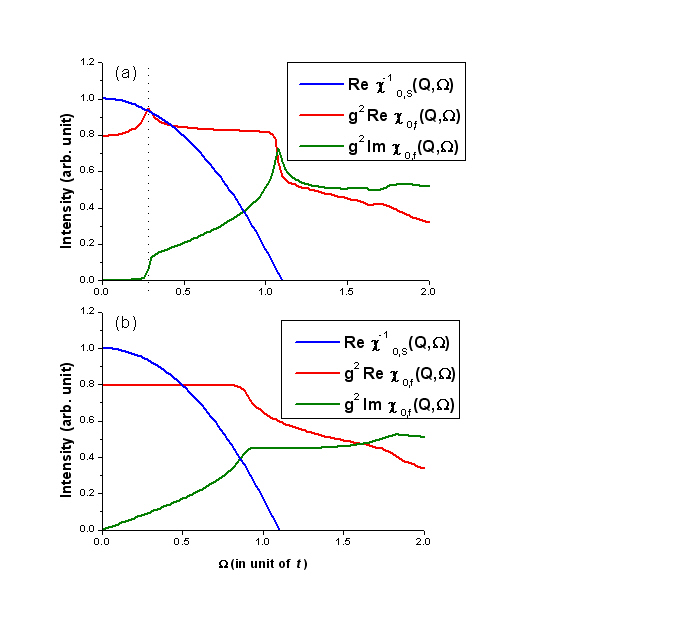}
\caption{(a) Plots of bare susceptibilities $Re
\chi^{-1}_{0,S}({\bf Q},\Omega)$, $g^2 Re \chi_{0,f}({\bf
Q},\Omega)$, and $g^2 Im \chi_{0,f}({\bf Q},\Omega)$, respectively
in superconducting state. Parameters are $\Delta_{SG}=1.1 t$,
$\Delta_0 = 0.2 t$, and $\lambda=0.8$. The vertical dashed line is
a guide to the eyes indicating the position of pole at
$\Omega=0.28 t$. (b) The same as (a) in normal state (ie.
$\Delta_0 =0$).} \label{fig2}
\end{figure}

The mechanism of forming the resonance mode in our model is
illustrated in Fig.2.  When the inverse of the dressed
susceptibilities of Eq.(3) and Eq.(4) crosses zero (which occurs
simultaneously in both susceptibilities), the dressed
susceptibilities develop a resonance mode, a bound state, or a
completely overdamped mode depending on the presence and the
strength of the imaginary part at the position of pole.
In Fig.2 we plot separately $Re \chi^{-1} _{0,S} ({\bf
Q},\Omega)$, $Re \chi_{0,f} ({\bf Q},\Omega)$, and $Im \chi_{0,f}
({\bf Q},\Omega)$ to make this point clear.
Fig.2(a) is the case of a superconducting state. By tuning the
coupling constant $g$ and the bare spin gap value $\Delta_{SG}$ as
chosen in Fig.1(a), the pole of $\chi_{f,S} ({\bf q =\bf
Q},\Omega)$ occurs at $\Omega_{res} \sim 0.28 t$. At this
frequency a small amount of $Im \chi_{0,f}$ still exists just
below the p-h excitation gap, making the pole a resonance mode.
Fig.2(b) shows the case of the normal state ($\Delta_0 =0$) with
the same parameters as in Fig.1(c). The position of pole occurs at
a little higher frequency ($\Omega \sim 0.5 t$) and there is
substantial amount of damping from $Im \chi_{0,f}$ making an
overdamped mode as shown in Fig.1(c).
With this mechanism of damping of the pole, even in the
superconducting state, a small variation of the coupling or the
superconducting gap, the intensity of the resonance mode can
vastly change. Finally, we would like to make a remark about the
unique character of the resonance mode in our model. The line of
$Re \chi^{-1} _{0,S} ({\bf Q},\Omega)$ in Fig.2 is not a simple
inverse of a static potential (for example, $\frac{1}{U}$ in a RPA
calculation of Hubbard model) but it carries its own dynamics and
spectral density. Therefore, the resonance mode formed by coupling
of two dynamic susceptibilities $\chi_{0,f}$ and $\chi_{0,S}$
should carry the spectral density from both the local spin wave
and the fermion particle-hole excitations.

To make a comparison of our calculations with experiments, it is
important to fix the energy scale of the model. The tight binding
band of Eq.(5) is widely studied to fit the ARPES data and the
estimate of $t$ varies from 150 meV to 300 meV depending on the
doping and different cuprate compounds \cite{band}. Our
calculation results are in good agreement with neutron experiments
in terms of energy scale if we choose $t \sim$ 150-180 meV. This
value of $t$ is somewhat low compared to the estimates from ARPES
experiments. One possible reason of this discrepancy is that the
extraction of $t$ value from ARPES is carried by fitting the whole
range of the quasiparticle dispersion. As a result the high energy
dispersion sets the overall energy scale and the value of $t$.
However, for calculations of the spin susceptibility the high
energy quasiparticle excitations are almost irrelevant and it is
the low energy quasiparticle excitations near Fermi level which
determine the structure of the low energy spin susceptibility.
With this reasoning it is quite possible that the effective $t$
value near FS is in fact much smaller due to a renormalization of
a strong correlation effect.
Also we use the chemical potential level $\mu=-0.8 t$ in our
calculations. This value of $\mu$ corresponds to the band filling
$n=0.503$ in our model, which is even higher than the half filling
($n=0.5$). In our phenomenological model the half filling has no
special meaning because we do not include any strong coupling
effect after the phenomenological Hamiltonian Eq.(1). On the other
hand the Fermi surface (FS) curvature is the single most important
parameter to control the degree of incommensurability (IC) and the
strength of the downward dispersion in the free fermion
susceptibility $\chi_{0,f}$  as shown in Fig.1(b). The curvature
can be tuned with $t^{'}$ and $\mu$ in our model. We decided to
fix $t^{'}=-0.4 t$ and tune the value of $\mu$ as a free
parameter. Using a lower value of $\mu$ (smaller band filling),
both the degree of the IC of the downward excitations and the p-h
excitations gap edge have tendency to increase. Using a more
realistic band with more tight binding parameters as $t^{''},
t^{'''}$, etc., it will be possible to choose more realistic band
fillings.

\begin{figure}
\noindent
\includegraphics[width=85mm]{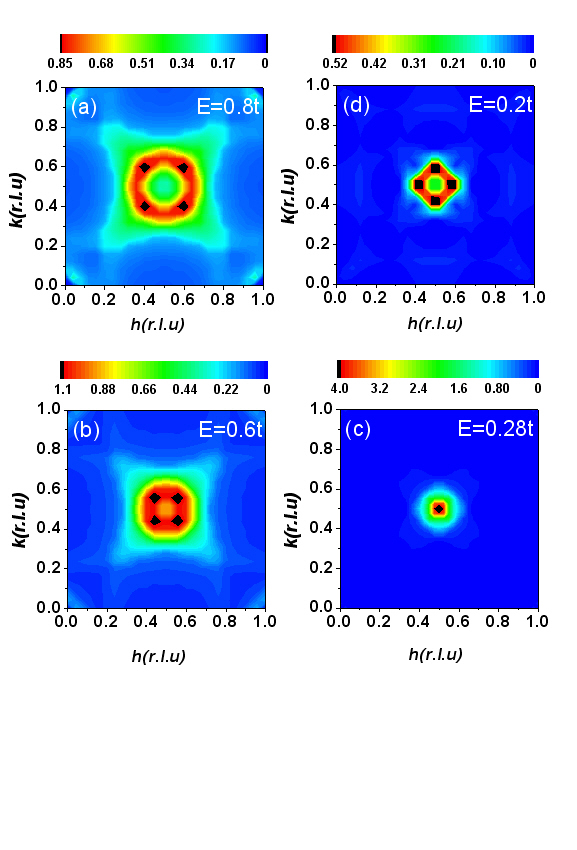}
\vspace{-3cm} \caption{Constant energy scans of $Im \chi_{f}({\bf
q},\Omega)$ at (a) $\Omega=0.8 t$, (b) $\Omega=0.6 t$, (c)
$\Omega=0.28 t$, (d) $\Omega=0.2 t$. Parameters are
$\Delta_{SG}=1.1 t$, $\Delta_0 = 0.2 t$, and $\lambda=0.8$.}
\label{fig3}
\end{figure}

Fig.3.(a-d) show the constant energy scans of $\chi_{f} ({\bf
q},\Omega)$ in the superconducting state for $\Omega=0.2t, 0.28t,
0.6t$, and $0.8t$, respectively.
Constant energy scans of neutron scattering data of YBCO
\cite{Hayden04} and LBCO \cite{Tranquada04} show peculiar patterns
of IC peak positions in $(q_x, q_y)$ momentum space at different
energy cuts. In particular, the 45 $\deg$ rotation of the patterns
from a low energy scan (below the resonance energy) to a high
energy scan draws special attention recently and several
theoretical explanations have been proposed
\cite{stripes-2,FL-theories-2}. Results of Fig.3(a-d) demonstrate
that the two component spin fermion model can consistently explain
this phenomena, too.
First, Fig.3(c) is the scan of $\chi_{f} ({\bf q},\Omega)$ at the
resonance energy, $\Omega=0.28 t$ with the same parameters as in
Fig.1(a). It shows a very intense peak at (1/2,1/2) indicating a
very sharp resonance not only in energy but also in momentum
space. Fig3(a) and Fig.3(b) are scans at higher energies than the
resonance energy and Fig.3(d) is a scan of lower energy cut. We
colored the highest intensity positions with black color to
emphasize the clear patterns. The lower energy scan (Fig.3(d))
shows the IC peaks at $(1/2 \pm \delta, 1/2)$ and $(1/2, 1/2 \pm
\delta)$ forming a diamond shape pattern. The higher energy scans
(Fig3(a) and Fig3(b)) show that the IC peak positions at $(1/2 \pm
\delta, 1/2 \pm \delta)$ and $(1/2 \pm \delta, 1/2 \mp \delta)$
forming a square shape pattern which has the symmetry of the 45
$\deg$ rotated from the low energy pattern. The results of Fig.3
excellently reproduce the observed patterns of the constant energy
scan data of neutron experiments reported in YBCO \cite{Hayden04}
and LBCO \cite{Tranquada04}.

In our model we can trace the origins of the patterns. The low
energy IC peaks and diamond shape pattern is basically a
reflection of the band structure and d-wave superconducting gap.
The high energy IC peaks and the square shape pattern has more
complicated origin. At and above the resonance energy the dressed
spin correlation $\chi_{f}$ is the result of a strong interplay
between the local spin correlation and the itinerant spin
correlation. Therefore the high energy scan pattern is determined
by a subtle interplay/competition between $\chi_{0,S}({\bf
q},{\Omega})$ and $\chi_{0,f}({\bf q},{\Omega})$.
The presence of IC peaks itself is the manifestation of the high
energy spin wave dispersion spanning from the AFM wave vector
${\bf Q}$; so the incommensurability increases with energy.
However, whether the pattern becomes a square or a diamond shape
has no universal mechanism.
We tested various combinations of parameters $t^{'}$, $\mu$,
$\Delta_{SG}$, $\Delta_{0}$ and $\lambda$. The low energy diamond
shape pattern is quite robust within our model. As to the patterns
of higher energy scans, although the square shape is the dominant
one, it is not absolutely robust; with different parameters the
diamond pattern often appears, too, at high energy scans.
Therefore, we think that the 45 $\deg$ rotation of the IC peak
patterns may not be a universal feature of high-T$_c$ cuprates; it
can change with doping and with different cuprate compounds.

In conclusion, we proposed a two-component spin fermion model to
explain the neutron scattering experiments in high-T$_c$ cuprates.
With the two spin degrees of freedom of the local spins and the
itinerant spins, our calculations of the spin correlations
reproduced the essential features of the experiments: the
hourglass dispersions, resonance mode, their changes in normal and
superconducting states, and the IC peak patterns of the constant
energy scans. With this success of the two-component spin fermion
model to describe the neutron scattering experiments, the pressing
question is now what the microscopic theory is for the
phenomenological two-component spin fermion model; specifically
how the local spin degree of freedom survives after doping from
the parent insulating cuprate compounds.

We thank A. Chubukov, I. Eremin, and B. Keimer for  discussions.
Y. B. was supported by the KOSEF through the CSCMR and the Grant
No. KRF-2005-070-C00044.

\end{document}